\documentclass[aps,prl,twocolumn,showpacs]{revtex4}

\usepackage{amssymb,amsmath,graphicx,graphics,color,amsfonts}

\newcommand{\im}[1]{\textrm{Im}\left\{  #1\right\}}

\newcommand{\de}{\partial}
\newcommand{\sech}[1]{\textrm{sech}\left(  #1\right)}

\newcommand{\eq}[2]{\begin{equation} \label{#1} #2 \end{equation}}

\newcommand{\etal}{{\em et al.}}

\newcommand{\cc}{\textrm{c.c.}}

\newcommand{\EE}{\mathbf{E}}

\newcommand{\rr}{\mathbf{r}}

\newcommand{\BB}{\mathbf{B}}
\newcommand{\DD}{\mathbf{D}}

\newcommand{\va}{\varepsilon}

\begin{document}

\title{Nonlinear wave dynamics in photonic time crystals}
\author{Fabio Biancalana}
\affiliation{Institute of Photonics and Quantum Sciences (IPaQS), School of Engineering and Physical Sciences, Heriot-Watt University, EH14 4AS Edinburgh, UK}

\begin{abstract}
Maxwell's wave equation in the presence of a cubic nonlinearity and a periodically time-varying refractive index (a photonic time crystal) is reduced, for spatially monochromatic waves, to a nonlinear Mathieu equation. Near the principal momentum gap this equation admits an autonomous two-dimensional reduction whose complete Hamiltonian phase portrait can be obtained analytically. We derive the two homoclinic separatrices corresponding to temporally localised momentum gap solitons, identify the nonlinear centres and the critical Hamiltonian value $H_c$, and calculate the point of maximum linear parametric gain. We then consider spatially localised pulses and show how the nucleation of multiple spatiotemporal gap solitons can produce a broad supercontinuum in momentum space; for stronger seeds, transient extreme nonlinear localisation can accompany an abrupt additional broadening of this momentum spectrum. These results establish a direct connection between Floquet amplification, nonlinear saturation, homoclinic dynamics, and momentum space spectral broadening in nonlinear photonic time crystals.
\end{abstract}

\maketitle

\section{Introduction}

The ability to control light by structuring matter in space is the central idea behind photonic crystals, metamaterials, and nonlinear waveguides. In these systems the electromagnetic frequency is conserved when the medium is time independent, while spatial momentum can be redistributed by spatial inhomogeneity. Time-varying media realise the complementary situation: if a material is spatially homogeneous but its constitutive parameters are modulated in time, spatial momentum remains the relevant conserved quantity, whereas electromagnetic energy and frequency need not be conserved \cite{morgenthaler,cassedy,fante,shv,mendonca_book,mendonca_basic,plansinis}. This exchange of the roles of time and space gives rise to temporal refraction, temporal reflection, parametric amplification, and momentum bandgaps, and has motivated the modern concept of photonic time crystals (PTCs) \cite{biancalana_pre,galiffi,caloz1,caloz2,lustig_optica,sharabi_disorder,dikopoltsev,lustig_fundamental,asgari}.

In a PTC the dielectric response is periodic in time and homogeneous, or nearly homogeneous, in space. The resulting band structure is therefore most naturally described at fixed wave vector rather than at fixed frequency. The gaps opened by the temporal modulation are momentum gaps, or $k$-gaps \cite{biancalana_pre}, in which Floquet modes with exponentially growing and decaying amplitudes coexist. This feature is fundamentally non-conservative: the time modulation acts as an external pump that can inject or extract electromagnetic energy from the field. Thus a PTC is not a conservative optical lattice obtained by simply exchanging space and time, but a driven electromagnetic medium in which amplification and nonlinear saturation are intrinsic features.

Nonlinear phenomena are not a peripheral extension of PTC theory; they constitute a substantial and rapidly developing part of the field. Because momentum gap amplification can drive the field amplitude upward, the material nonlinearity is naturally intertwined with the long time dynamics. Existing studies have addressed, among other effects, superluminal $k$-gap solitons, Kerr-induced symmetry breaking and spatiotemporal pattern formation, nonlinear frequency conversion, and nonlinear spatio-spectral fission in strongly time-varying media \cite{biancalana_gap,segevprl,kiselev,konforty,jaffray}; broader perspectives are given in Refs.~\cite{galiffi,asgari}.

The experimental relevance of this physics is now rapidly increasing. At visible and infrared frequencies, the principal obstacle to a genuine PTC is no longer the absence of candidate mechanisms, but the simultaneous material requirements of a sizeable refractive-index modulation depth, subcycle switching, sufficiently low loss, and a recovery fast enough to repeat the modulation periodically. These constraints, and the material platforms that may satisfy them, have been analysed in detail from a material science perspective \cite{saha_materials}.
Transparent conducting oxides (TCOs) operating near their epsilon-near-zero (ENZ) point have emerged as particularly promising systems because modest changes in their electronic response can translate into very large changes of refractive index. Single-cycle time refraction has already been observed in TCOs \cite{lustig_singlecycle}. Most recently, Segal \etal\ demonstrated in doped CdO an index change of approximately $\Delta n\simeq-0.3$ occurring within about $0.77$ optical cycles for a $4~\mu$m probe, driven by a $10$~fs optical modulator pulse \cite{segal_subcycle}. At microwave and terahertz frequencies, genuine PTC behaviour has now also been demonstrated experimentally: Xiong \etal\ observed $k$-gap amplification and temporal topology in a dynamically modulated transmission line PTC \cite{xiong_ptc}, while Guo \etal\ reported an all-optical plasmonic metamaterial PTC operating at terahertz frequencies with coherent subcycle modulation and a measured transition into the PTC regime \cite{guo_ptc}. A complementary route is to exploit material or structural resonances, which can strongly enlarge the $k$-gap for much smaller modulation depths \cite{wang_resonant}.

The present work develops a unified description of nonlinear dynamics near the first momentum gap of a PTC. For spatially monochromatic fields, we reduce the problem to a nonlinear Mathieu equation and obtain the temporal gap soliton solution throughout the first instability tongue, together with its Hamiltonian phase portrait and the point of maximum Floquet gain. This extends previous treatments by providing a complete analytical characterisation of the homoclinic dynamics within the reduced model. We then retain the spatial dependence through a forward/backward coupled-mode description. Starting from a single spatially localised forward Gaussian pulse, we show that the interplay of temporal Bragg amplification and Kerr nonlinearity produces multiple spatiotemporal gap solitons and a broad redistribution of momentum, leading to {\em momentum supercontinuum generation}. For stronger excitation, the same dynamics can produce transient extreme localisation accompanied by an abrupt further expansion of the momentum spectrum.

\section{Governing equations}

In order to study the temporal dynamics of a homogeneous material whose refractive index changes periodically in time (also called {\em photonic time crystal}, or PTC), we start from Maxwell's equations
\begin{eqnarray}
\nabla\times\BB &=& \frac{1}{c}\de_{t}\DD, \label{max1}\\
\nabla\times\EE &=& -\frac{1}{c}\de_{t}\BB, \label{max2}
\end{eqnarray}
complemented with the nonlinear relation between electric and displacement fields: $\EE=\DD/\va(t)-\chi^{(3)}\DD^{3}/\va^{4}(t)$. It is customary to work with $\DD$ instead of $\EE$, since the displacement field is continuous across time interfaces, and also because this simplifies the following algebra considerably.

The wave equation for the displacement field $\DD(\rr,t)$ in a cubic nonlinear dielectric medium is thus:
\eq{eq1}{\frac{1}{c^2}\de_{t}^{2}\DD-\nabla^2\left(\frac{\DD}{\va(t)}-\frac{\chi^{(3)}}{\va(t)^4}\DD^{3}\right)=0.} 
In equations (\ref{max1},\ref{max2},\ref{eq1}), $\va(t+T_{\rm m})=\va(t)$ is the dielectric function oscillating with modulation period $T_{\rm m}$, $\chi^{(3)}$ is the third-order nonlinear susceptibility (expressed in units of m$^2$/V$^2$, and assumed to be a constant), and $c$ is the speed of light in vacuum. Equations (\ref{max1},\ref{max2},\ref{eq1}) are written in the Heaviside-Lorentz units, where all fields (electric, magnetic, polarization, displacement, etc...) are expressed in units of V/m.

We now assume a standard periodic variation of $\va$, namely $\va(t)\equiv\va_{r}f(t)=\va_{r}\left[1+\Delta \cos(\Omega t)\right]$, where $\va_{r}$ is the dimensionless base relative dielectric constant (around which the periodic oscillation occurs), $\Delta$ is the depth of the dielectric constant modulation, and $\Omega\equiv 2\pi/T_{\rm m}$ is the modulation frequency. Specialising to one spatial dimension ($x$), and assuming that the displacement field is spatially monochromatic with wavenumber $k$, we write $D(x,t)\equiv\frac{1}{2}\left(e^{ikx}+e^{-ikx}\right)A(t)$, and we obtain, grouping for either $e^{ikx}$ or $e^{-ikx}$ and neglecting all spatial third-harmonic generation terms oscillating with $e^{\pm 3ikx}$:

\eq{eq2}{\ddot{\mathcal{A}}+\left(\frac{kct_{0}}{n_{r}}\right)^{2}\frac{1}{f}\left[1-\frac{3\chi^{(3)}A_{\rm s}^2}{4 n_{r}^6f^3}\mathcal{A}^2\right]\mathcal{A}=0,} where we have rescaled time as $t\equiv t_{0}\tau$ (stressing the necessity of temporal interfaces varying at a scale comparable to the optical cycle \cite{lustig_fundamental,segevprl}) and amplitude as $A\equiv A_{\rm s}\mathcal{A}$, and where $f(\tau)\equiv 1+\Delta\cos(\Omega t_{0}\tau)$ and $n_{r}\equiv\sqrt{\va_{r}}$ is the background average refractive index of the medium. The double dot in Eq. (\ref{eq2}) refers to double derivative in $\tau$.

We further conveniently fix the temporal scale to be $t_{0}\equiv 2/\Omega$ and the amplitude scale to be $A_{\rm s}\equiv[(4n_{r}^6)/(3\chi^{(3)})]^{1/2}$, and with the definition of the electromagnetic frequency $\omega_{0}\equiv kc/n_{r}$ and of the parameter $a\equiv (2\omega_{0}/\Omega)^2$ we obtain
\eq{eq3}{\ddot{\mathcal{A}}+\frac{a}{f}\left[1-\frac{\mathcal{A}^2}{f^{3}}\right]\mathcal{A}=0.}
Equation (\ref{eq3}) is a nonlinear Hill-type equation that is at the core of our investigation of spatially monochromatic modes in cubic periodically time-varying media. In the following we assume an instantaneous self-focusing Kerr response, $\chi^{(3)}>0$.

\section{Nonlinear Mathieu equation}

For a shallow modulation depth $\Delta$, we expand the linear factor $f^{-1}$ to first order and retain its resonant contribution at the first temporal Bragg condition. In Eq.~(\ref{eq3}) the cubic term is proportional to $f^{-4}\mathcal A^3$; within the same shallow modulation first-gap reduction we neglect the residual periodic modulation of this nonlinear prefactor and set $f^{-4}\rightarrow1$. This approximation can be checked {\em a posteriori} against the full nonlinear Hill equation and is consistent with the coupled-mode approximations used in previous nonlinear PTC treatments, including Refs.~\cite{segevprl,kiselev}. This leads us to the study of the following {\em nonlinear Mathieu equation} (NME) for the amplitude, which we write in the standard form conventionally used in the literature ($q\equiv a\Delta /2$):
\eq{eq4}{\ddot{\mathcal{A}}+\left[a-2q\cos(2\tau)\right]\mathcal{A}-a\mathcal{A}^3=0.}
One important feature of Eq. (\ref{eq4}), that distinguishes it from the standard formulation of Mathieu's equation found in textbooks and in the literature, is that $a$ and $q$ are not independent in our nonlinear optical case. This constraint makes the optical problem different from many mathematical treatments of weakly nonlinear Mathieu equations, where the nonlinear coefficient is an independent perturbative parameter.

For the harmonic modulations of the refractive index considered here, the Floquet theorem must be used, and the linear (undamped) version of Eq. (\ref{eq4}) possesses an infinite number of parametric instability regions in the ($a,q$) parameter space (called {\em Arnold's tongues}), emanating from the points $a=n^2$ (the temporal equivalent of the Bragg condition that is found in spatially periodic media), where $n\in\mathbb{Z}$. In our physical units this would correspond to $\omega_{0}=n\Omega/2$. Inside these regions, determined by the Floquet criterion, the field amplitude $\mathcal{A}(\tau)$ of the linearized version of Eq. (\ref{eq4}) will grow indefinitely with time. 

The Arnold tongues become increasingly narrower as the Bragg order $n$ increases. We can now apply the known published results on the bandgap range for each Bragg order. If the parameters $a$ and $q$ were independent, the first ($n=1$) Bragg gap exists in the range $1-q<a<1+q$ [up to $\mathcal{O}(q^2)$], while the second Bragg gap ($n=2$) exists in the range $4-q^2/12<a<4+5q^{2}/12$  [up to $\mathcal{O}(q^4)$]. Higher-order gap ranges can be found in \cite{mathieubook}. In our case, however, the relation $q\equiv a\Delta /2$ implies that the stability tongues of the linearised version of Eq. (\ref{eq4}) change, and a simple calculation for the first two Bragg orders results in the following ranges:

\begin{eqnarray}
n=1:\ (1+\Delta/2)^{-1}<&a&<(1-\Delta/2)^{-1} , \label{bragg1}\\
n=2:\ 8\left[\frac{\sqrt{9+3\Delta^2}-3}{\Delta^2}\right]<&a&<\frac{24-8\sqrt{9-15\Delta^2}}{5\Delta^2} , \label{bragg2}
\end{eqnarray}

The expressions become increasingly more complicated for higher Bragg orders, while the higher-order instability regions become rapidly narrower for shallow modulation. In the Mathieu parameter $a$, the first gap width is $\delta a_{1}=4\Delta/(4-\Delta^2)\simeq\Delta$, whereas the second gap width is $\delta a_{2}\simeq2\Delta^2$. Since $\omega_0=(\Omega/2)\sqrt a$, these correspond, to leading order, to widths $\delta\omega_{0,1}\simeq\Delta\Omega/4$ and $\delta\omega_{0,2}\simeq\Delta^2\Omega/4$, respectively. Here $\omega_0=kc/n_r$ is the bare frequency parameter used to label the conserved momentum $k$; these widths should therefore not be interpreted as ordinary gaps in a conserved optical frequency. We focus below on the first gap ($n=1$), while higher-order gaps admit analogous but progressively less accessible reductions.

\section{Momentum gap solitons from nonlinear Mathieu's equation}
Starting from Eq. (\ref{eq4}), we can define a complex temporal envelope field $M(\tau)$ such that $\mathcal{A}(\tau)\equiv M(\tau)e^{-i\tau}+M^{*}(\tau)e^{+i\tau}$. This amounts to restricting our study to the $n=1$ Bragg gap, i.e. the physically more relevant momentum gap. Substituting into Eq. (\ref{eq4}), we obtain 
\eq{eqforM}{i\dot{M}+\frac{1}{2}(1-a)M+\frac{1}{4}a\Delta M^{*}+\frac{3}{2}a|M|^{2}M=0,}
and separating the real and imaginary parts of $M(\tau)\equiv u(\tau)+i v(\tau)$ in Eq. (\ref{eqforM}), we obtain the following nonlinear dynamical system:
\begin{eqnarray}
\dot{u}=\frac{1}{2}\left[\left(1+\frac{1}{2}\Delta\right)a-1\right]v-\frac{3}{2}a\left(u^2+v^2\right)v, \label{coup1} \\
\dot{v}=-\frac{1}{2}\left[\left(1-\frac{1}{2}\Delta\right)a-1\right]u+\frac{3}{2}a\left(u^2+v^2\right)u. \label{coup2}
\end{eqnarray}
In the derivation of Eqs. (\ref{coup1},\ref{coup2}) we have again neglected the third harmonic terms oscillating like $e^{\pm 3i\tau}$. It is evident from the structure of Eqs. (\ref{coup1},\ref{coup2}) that the edges of the momentum gap are located exactly as described by Eq. (\ref{bragg1}). Note that similar equations have been recently used in order to study phase transitions and spatiotemporal pattern formation in PTCs in presence of losses, in the very interesting work \cite{kiselev}. It is worth noting that if one desires to study a different momentum bandgap, or more than one bandgap, an expansion $\mathcal{A}(\tau)\equiv \sum_{m=1}^{N}\left(M_{m}(\tau)e^{-im\tau}+M_{m}^{*}(\tau)e^{+im\tau}\right)$ is necessary, which would produce $N$ pairs of coupled equations for the different envelopes describing the field in the $m$-th bandgap.

Upon changing the variables to polar coordinate, $u\equiv r\cos(\theta)$ and $v\equiv r\sin(\theta)$, Eqs. (\ref{coup1},\ref{coup2}) become:
\begin{eqnarray}
\dot{\rho}&=&\frac{a\Delta}{2}\rho\sin(2\theta), \label{pol1} \\
\dot{\theta}&=&\frac{1}{2}\left[1-a+\frac{a\Delta}{2}\cos(2\theta)\right]+\frac{3}{2}a\rho, \label{pol2}
\end{eqnarray}
where we have defined $\rho\equiv r^{2}$.

The phase portrait is that of a two-dimensional autonomous nonlinear system with a Hamiltonian separatrix structure \cite{strogatz}.

From Eqs. (\ref{pol1},\ref{pol2}) we deduce that
\eq{alg1}{\frac{d\rho}{d\theta}=\frac{a\Delta\rho\sin(2\theta)}{\left[1-a+\frac{a\Delta}{2}\cos(2\theta)\right]+3a\rho},}
which can be integrated by specifying the appropriate boundary conditions.

The position of the fixed points of Eqs. (\ref{coup1},\ref{coup2}) for those values of $a$ strictly inside the momentum gap are $(u_{0},v_{0})=(0,0)$ (saddle point), $(u_{\pm},v_{\pm})=\left(0,\pm\sqrt{\frac{(2+\Delta)a-2}{6a}}\right)$ (nonlinear centres).

Focusing on the saddle point $(u_{0},v_{0})$ will give us the homoclinic orbit, related to the momentum gap soliton. The boundary conditions for the temporally localised solitonic solutions are $\rho(\pm\infty)=0$.
The maximum amplitude reached during the trajectory on the homoclinic orbit can be deduced by using Eq. (\ref{alg1}): $\rho_{max}=\Delta/3+2(a-1)/(3a)$. The initial angle $\theta_{i}$ of the trajectory from the saddle point is given by setting $\rho=0$ again in Eq. (\ref{alg1}), and is given by:
\eq{inang}{\theta_{i}\equiv\theta(\tau=-\infty)=\frac{1}{2}\arccos\left[\frac{2(a-1)}{a\Delta}\right].} This angle corresponds to $\tau=-\infty$, while at $\tau=+\infty$ we have $\theta_{f}=\pi-\theta_{i}$, since the trajectory must be symmetric with respect to $\tau=0$. This is the point where the amplitude of the wave reaches its maximum $\rho(\tau=0)=\rho_{max}$.

We have found the general analytical solution of the dynamical system Eqs. (\ref{pol1},\ref{pol2}) for the momentum gap soliton with values of $a$ inside the gap:
\begin{widetext}
\eq{rho1}{\rho(\tau)=\frac{1}{3a}\frac{a\left[8+a(\Delta^{2}-4)\right]-4}{2(1-a)+a\Delta\cosh\left(\frac{1}{2}a\sqrt{\Delta^{2}-\frac{4(a-1)^{2}}{a^{2}}}\tau\right)}}
\eq{theta1}{\theta(\tau)=\frac{1}{2}\arccos\left[\xi(\tau)\right]\Theta(-\tau)-\left\{\frac{1}{2}\arccos\left[\xi(\tau)\right]-\pi\right\}\Theta(\tau) } 
\end{widetext}
where we have defined the quantity $\xi(\tau)\equiv\frac{2(a-1)-3a\rho(\tau)}{a\Delta}$, and $\Theta$ is the Heaviside function, which ensures that $\theta(\tau)$ is continuous, its derivative is also continuous, and $\theta_{i}+\theta_{f}=\pi$ for the homoclinic trajectory. The corresponding spatial gap soliton result in Ref. \cite{coste} was expressed in terms of $\tan\theta$, which determines the phase only modulo $\pi$; reconstructing a continuous complex envelope therefore requires an implicit branch continuation across the soliton centre. Equation~(\ref{theta1}) makes this branch prescription explicit. The explicit solution Eqs. (\ref{rho1},\ref{theta1}), derived via a nonlinear Mathieu equation analysis, is the first main result of this paper. Given a solution $(u,v)$, also $(-u,-v)$ will be a solution, due to the symmetry $u\rightarrow -u$ and $v\rightarrow -v$ of Eqs. (\ref{coup1},\ref{coup2}).

The specific case of momentum gap solitons at exactly $a=1$ (typically very close to the bandgap centre, which is located at $a=a_{0}\equiv4/(4-\Delta^2)$), found with the above method, is given by:
\begin{eqnarray}
u(\tau)&=&\pm\sqrt{\frac{\Delta}{3}\sech{\frac{\Delta\tau}{2}}}\left(\frac{1-e^{\Delta\tau/2}}{\sqrt{2+2e^{\Delta\tau}}}\right), \\
v(\tau)&=&\pm\sqrt{\frac{\Delta}{3}\sech{\frac{\Delta\tau}{2}}}\sqrt{\frac{1+\sech{\frac{\Delta\tau}{2}}}{2}},
\end{eqnarray}
and the angle $\theta(\tau)=\left(\pi+Gd(\Delta\tau/2)\right)/2$, where $Gd(x)\equiv 2\arctan(\tanh(x/2))$ is the Gudermannian function. The $+$ sign selects the homoclinic orbit in the upper half-plane, while the $-$ sign selects its symmetry-related counterpart in the lower half-plane. In the exact deterministic system each homoclinic loop is an invariant trajectory, so a solution cannot switch between the two loops; the possible role of fluctuations near the saddle is discussed below.

At the upper bandgap edge, when $a$ tends to the value $(1-\Delta/2)^{-1}$, the limiting value of solution Eq. (\ref{rho1}) is found to be the Lorentzian
\eq{lor1}{\rho_{up}(\tau)=\frac{2\Delta(\Delta-2)^{2}}{3\left[4(1-\Delta)+\Delta^{2}(1+\tau^{2})\right]},}
with maximum peak value equal to $\rho_{max}=2\Delta/3$ and temporal width $\tau_{0}=|(2-\Delta)/\Delta|$.
In close proximity to the lower bandgap edge, when $a$ tends to the value $(1+\Delta/2)^{-1}+\epsilon$, with $\epsilon$ positive and small, the gap soliton tends to a constant $\rho_{low}(\tau)\simeq (\Delta+2)^{2}\epsilon/6$, with a vanishing amplitude and an infinite temporal width for $\epsilon=0$, i.e. when exactly touching the lower bandgap edge.

Physically, the temporal gap soliton solution described in this section is a spatially extended plane-wave mode whose envelope is localised in time. In the ideal spatially monochromatic limit, the field intensity rises from an asymptotically vanishing value, forms a finite solitonic burst, and then returns to zero at every position in the homogeneous medium. It can therefore be viewed as a temporal flash of light, or a `photonic lighthouse', rather than as a conventional pulse localised and propagating in space. This interpretation refers to the slowly varying envelope; the underlying carrier retains its plane-wave spatial dependence. The deterministic homoclinic trajectory approaches the saddle point only asymptotically and, by uniqueness of the equations of motion, remains on the same loop. Nevertheless, sufficiently close to the saddle point the field amplitude is exponentially small, so weak noise, disorder, or numerical perturbations can dominate the subsequent departure from its neighbourhood. The sign of the perturbation projected onto the unstable eigendirection then selects the upper or lower homoclinic loop, corresponding to the symmetry-related solutions $M$ and $-M$. Successive bursts in a weakly perturbed or stochastic system may therefore switch between the two loops; for statistically unbiased fluctuations the two choices are symmetry-equivalent, although any systematic phase or amplitude bias can favour one of them. A closely related stochastic switching between the two halves of the phase portrait was observed numerically by Coste and Peyraud for spatial gap solitons in the full periodically modulated system, where successive localised peaks acquired random signs \cite{coste}.

\section{Hamiltonian phase portrait and maximum linear gain}

The autonomous system in Eqs.~(\ref{coup1}) and (\ref{coup2}) admits a particularly transparent Hamiltonian formulation. It is convenient to introduce the two coefficients
\begin{eqnarray}
 c_1&\equiv&\frac{1}{2}\left[\left(1+\frac{\Delta}{2}\right)a-1\right], \label{c1}\\
 c_2&\equiv&\frac{1}{2}\left[1-\left(1-\frac{\Delta}{2}\right)a\right].
 \label{c2}
\end{eqnarray}
The equations of motion may then be written as
\begin{eqnarray}
 \dot u&=&c_1v-\frac{3a}{2}(u^2+v^2)v,\label{ham_system1} \\
 \dot v&=&c_2u+\frac{3a}{2}(u^2+v^2)u,
 \label{ham_system2}
\end{eqnarray}
and follow from the conserved Hamiltonian
\eq{ham_phase}{H(u,v)=\frac{c_1}{2}v^2-\frac{c_2}{2}u^2-\frac{3a}{8}(u^2+v^2)^2,}
through $\dot u=\partial H/\partial v$ and $\dot v=-\partial H/\partial u$. Direct substitution gives $dH/d\tau=0$.

To determine the character of the trivial fixed point, we linearise the vector field in Eqs.~(\ref{ham_system1},\ref{ham_system2}) around $(u,v)=(0,0)$. The corresponding Jacobian matrix is
\eq{jacobian_origin}{J_0=\left.\frac{\partial(\dot u,\dot v)}{\partial(u,v)}\right|_{(0,0)}=
\begin{pmatrix}
0 & c_1\\
c_2 & 0
\end{pmatrix}.}
Its characteristic equation is $\det(J_0-\lambda I)=\lambda^2-c_1c_2=0$, and therefore
\eq{saddle_eigenvalues}{\lambda_{\pm}=\pm\sqrt{c_1c_2}.}
For values of $a$ strictly inside the first momentum gap, see Eq.~(\ref{bragg1}), both $c_1$ and $c_2$ are positive. The two eigenvalues are consequently real, nonzero, and of opposite sign, proving that the origin is a hyperbolic saddle. The associated unstable and stable eigendirections may be chosen as
\eq{saddle_eigenvectors}{\mathbf{e}_{+}=\begin{pmatrix}\sqrt{c_1}\\ \sqrt{c_2}\end{pmatrix},\qquad
\mathbf{e}_{-}=\begin{pmatrix}\sqrt{c_1}\\-\sqrt{c_2}\end{pmatrix},}
respectively. The two homoclinic soliton loops approach the origin tangentially to these invariant directions as $\tau\to\pm\infty$. At either band edge $c_1c_2=0$, so the fixed point becomes nonhyperbolic and the exponential growth rate vanishes.

The remaining fixed points are the two symmetry-related nonlinear centres
\eq{nonlinear_centres}{(u_c,v_c)=\left(0,\,\pm\sqrt{\frac{2c_1}{3a}}\right)=\left(0,\,\pm\sqrt{\frac{(2+\Delta)a-2}{6a}}\right).}
The Hamiltonian takes the same value at both centres. This critical value is
\begin{eqnarray}
H_c&\equiv&H(0,v_c)
=\frac{c_1}{2}\frac{2c_1}{3a}
-\frac{3a}{8}\left(\frac{2c_1}{3a}\right)^2
=\frac{c_1^2}{6a} \nonumber\\
&=&\frac{\left[(2+\Delta)a-2\right]^2}{96a}.
\label{Hc}
\end{eqnarray}
Thus $H_c$ is the maximum Hamiltonian value attained within either homoclinic lobe. At the maximum gain value of $a$ derived below in this section, it becomes
\eq{Hc_at_gain}{H_c(a_0)=\frac{\Delta^2(2+\Delta)}{96(2-\Delta)},}
for $0<\Delta<2$.

The global topology is most easily understood in polar variables, $u=\sqrt{\rho}\cos\theta$ and $v=\sqrt{\rho}\sin\theta$, for which
\eq{Hpolar}{H(\rho,\theta)=\frac{\rho}{2}A(\theta)-\frac{3a}{8}\rho^2,\qquad A(\theta)=c_1\sin^2\theta-c_2\cos^2\theta.}
The phase portrait contains three classes of level sets. For $0<H<H_c$, the level set consists of two disconnected periodic orbits, one surrounding each nonlinear centre. At $H=H_c$ each periodic family contracts to its centre. At $H=0$ the two periodic families terminate on two homoclinic separatrices that leave the saddle and return to it as $\tau\rightarrow\pm\infty$. These are precisely the two symmetry-related temporal gap solitons $M(\tau)$ and $-M(\tau)$ derived in Eqs.~(\ref{rho1}) and (\ref{theta1}). Their polar equation is
\eq{rho_homoclinic}{\rho_{\rm h}(\theta)=\frac{4}{3a}A(\theta),\qquad A(\theta)\geq0,}
with asymptotic angles $\theta_i$ and $\pi-\theta_i$ given by Eq.~(\ref{inang}). Finally, for $H=h<0$ there is one outer periodic orbit enclosing both lobes and the saddle. Indeed, solving Eq.~(\ref{Hpolar}) for the positive radial root gives
\eq{rho_outer}{\rho_h(\theta)=\frac{2}{3a}\left[A(\theta)+\sqrt{A^2(\theta)-6ah}\right],\qquad h<0,}
which is positive and $\pi$-periodic for every $\theta$. Therefore the negative Hamiltonian trajectories are closed outer orbits, not unbounded trajectories.

\begin{figure*}[t]
\centering
\includegraphics[width=0.7\linewidth]{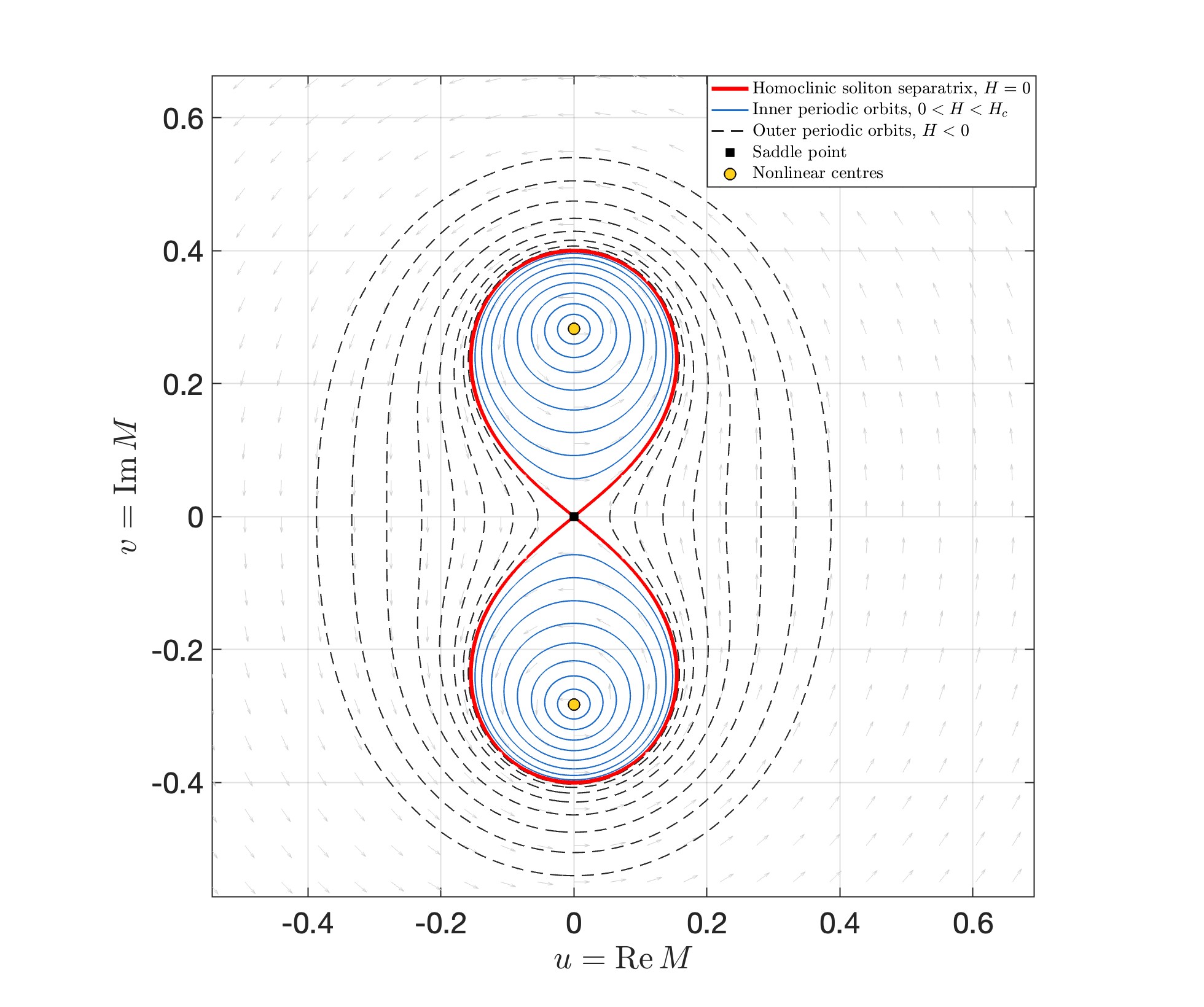}
\caption{Hamiltonian phase portrait of the autonomous first gap reduction for $\Delta=0.4$ and $a=a_0\equiv4/(4-\Delta^2)\simeq 1.041\bar{6}$, the point of maximum linear gain. The black square at the origin is the hyperbolic saddle, and the yellow circles are the two nonlinear centres of Eq.~(\ref{nonlinear_centres}). Blue solid contours are the two disconnected families of periodic trajectories with $0<H<H_c$, each surrounding one centre. The thick red `figure of eight' is the exact zero-Hamiltonian separatrix; its upper and lower lobes are the symmetry-related homoclinic temporal gap solitons $M(\tau)$ and $-M(\tau)$. Black dashed contours are the closed outer periodic trajectories with $H<0$, which enclose both centres and the saddle point.}
\label{fig:phase_portrait}
\end{figure*}

Figure~\ref{fig:phase_portrait} summarises this phase-space structure at the representative maximum gain point $\Delta=0.4$ and $a=a_0=4/(4-\Delta^2)$. The two red branches of the $H=0$ `figure of eight' are the symmetry-related homoclinic solitons, while the blue contours inside each lobe describe nonlinear periodic oscillations about the two centres. The dashed contours outside the separatrix are also closed periodic trajectories; thus the separatrix distinguishes the two independent inner oscillation families from a single outer family that encloses both centres and the saddle.

We now determine analytically the point of maximum linear gain. Near the saddle, the nonlinear terms in Eqs.~(\ref{ham_system1},\ref{ham_system2}) may be neglected, yielding $\ddot u=c_1c_2u$ and $\ddot v=c_1c_2v$. The positive exponential growth rate is therefore
\eq{lambda_general}{\lambda(a)=\sqrt{c_1c_2},}
or, explicitly,
\eq{lambda_squared}{\lambda^2(a)=\frac{1}{4}\left\{2a-\left(1-\frac{\Delta^2}{4}\right)a^2-1\right\}.}
This is a concave quadratic function of $a$. Since $\lambda\geq0$, maximising $\lambda$ is equivalent to maximising $\lambda^2$. Differentiation gives
\begin{eqnarray}
\frac{d\lambda^2}{da}&=&\frac{1}{2}\left[1-\left(1-\frac{\Delta^2}{4}\right)a\right], \label{gain_derivative}\\
\frac{d^2\lambda^2}{da^2}&=&-\frac{1}{2}\left(1-\frac{\Delta^2}{4}\right)<0.
\label{gain_curvature}
\end{eqnarray}
Consequently, for $|\Delta|<2$, the unique maximum occurs at
\eq{a_max_gain}{a=a_0=\frac{1}{1-\Delta^2/4}=\frac{4}{4-\Delta^2},} which is the expression we have used throughout this paper.
This value is exactly the arithmetic midpoint, in the variable $a$, of the two first-gap edges,
\eq{gap_midpoint}{a_0=\frac{1}{2}\left[\frac{1}{1+\Delta/2}+\frac{1}{1-\Delta/2}\right].}
The corresponding maximum gain is
\eq{lambda_max}{\lambda_0=\frac{|\Delta|}{2\sqrt{4-\Delta^2}}\simeq\frac{|\Delta|}{4},}
where the final expression applies for $|\Delta|\ll1$. The statement that $a_0$ is the maximum gain point is exact within the autonomous first gap reduction. For the full nonlinear Mathieu equation at larger modulation depths, higher-order Floquet corrections can shift the exact maximum slightly.

\section{Spatiotemporal gap solitons and momentum supercontinuum generation}

The spatially monochromatic analysis isolates the temporal nonlinear states associated with a single wave vector. A realistic excitation, however, has finite spatial width and therefore occupies a range of momenta. The appropriate extension is obtained by retaining the resonant forward and backward components close to the first temporal Bragg condition. We set $\Omega=2\omega_0$ (this corresponds to $a=1$ in the notation of the preceding sections), $k_0=n_r\omega_0/c$, and write, in one spatial dimension, 
\eq{FB_ansatz}{D(z,t)=\frac{1}{2}\left[F_{\rm d}(z,t)e^{i(k_0z-\omega_0t)}
+B_{\rm d}(z,t)e^{i(k_0z+\omega_0t)}+\cc\right],}
where $F_{\rm d}$ and $B_{\rm d}$ are slowly varying compared with $\omega_0^{-1}$ and $k_0^{-1}$. Substitution in Eq.~(\ref{eq1}), followed by the slowly varying envelope approximation and a rotating wave projection onto the two resonant carriers, gives
\begin{eqnarray}
i(\de_t+v\de_z)F_{\rm d}+\kappa B_{\rm d}
+\gamma_{\rm d}\left(|F_{\rm d}|^2+2|B_{\rm d}|^2\right)F_{\rm d}&=&0,\nonumber\\
i(\de_t-v\de_z)B_{\rm d}-\kappa F_{\rm d}
-\gamma_{\rm d}\left(|B_{\rm d}|^2+2|F_{\rm d}|^2\right)B_{\rm d}&=&0,
\label{dimensional_coupled}
\end{eqnarray}
where
\eq{dimensional_coefficients}{v\equiv\frac{c}{n_r},\qquad 
\kappa\equiv\frac{\Delta\cdot\omega_0}{4},\qquad
\gamma_{\rm d}\equiv\frac{3\chi^{(3)}\omega_0}{8n_r^6}.}
The coefficient $\kappa$ is the temporal Bragg coupling produced by the harmonic modulation, while the factor of two in the cross-phase terms comes from the resonant part of $D^3$. Nonresonant carrier components, including third harmonics, have been neglected.

Finally, we scale time, space, and field amplitude as
\eq{FB_scaling}{t\rightarrow \kappa t,\qquad z\rightarrow \frac{\kappa}{v}z,\qquad
F\equiv\sqrt{\frac{\gamma_{\rm d}}{\kappa}}F_{\rm d},\qquad
B\equiv\sqrt{\frac{\gamma_{\rm d}}{\kappa}}B_{\rm d}.}
Equivalently, the dimensional displacement field scale is
\eq{Ds_scale}{D_s\equiv\sqrt{\frac{\kappa}{\gamma_{\rm d}}}
=n_r^3\sqrt{\frac{2\Delta}{3\chi^{(3)}}}.}
After this normalization, and dropping the distinction between scaled and unscaled coordinates, Eq.~(\ref{dimensional_coupled}) becomes \cite{biancalana_gap,segevprl}:
\begin{eqnarray}
i(\de_{t}+\de_{z})F+B+(|F|^2+2|B|^2)F&=&0, \nonumber\\
i(\de_{t}-\de_{z})B-F-(|B|^2+2|F|^2)B&=&0.
\label{coupled1}
\end{eqnarray}
All wavenumbers appearing below are correspondingly understood in units of $\kappa/v$, and frequencies in units of $\kappa$; thus the carrier quantities become $k_0\rightarrow vk_0/\kappa=4/\Delta$ and $\omega_0\rightarrow\omega_0/\kappa=4/\Delta$. For future use, we also introduce the dimensionless momentum detuning $\Delta k\equiv k-k_0$.

For spatially localised fields, the two relevant integral quantities in this reduced model are
\begin{eqnarray}
\mathcal{E}(t)&\equiv&\int_{-\infty}^{+\infty}\left(|F|^{2}+|B|^{2}\right)dz, \label{em_energy}\\
\mathcal{P}(t)&\equiv&\int_{-\infty}^{+\infty}\left(|F|^{2}-|B|^{2}\right)dz.
\label{em_momentum}
\end{eqnarray}
A direct calculation from Eq.~(\ref{coupled1}) gives
\eq{em_balance}{\frac{d\mathcal{E}}{dt}=4\int_{-\infty}^{+\infty}\im{FB^{*}}dz,\qquad \frac{d\mathcal{P}}{dt}=0.}
Thus the temporal modulation can exchange energy with the optical field, whereas spatial translation invariance protects the total momentum $\mathcal P$. These balance laws are also useful numerics: a converged simulation should preserve $\mathcal P$ to the accuracy of the integration scheme and should reproduce the nonzero energy balance term in Eq.~(\ref{em_balance}).

The coupled-mode equations support nonlinear states continuously connected to the homoclinic temporal gap solitons discussed in previous sections, which we call {\em spatiotemporal gap solitons}. In fact, Eq.~(\ref{coupled1}) admits the exact travelling family
\cite{biancalana_gap}
\begin{eqnarray}
F(z,t)&=&A_F\sqrt{\sech{2\zeta}}\,
\exp\left\{i\left[\phi_0+\nu_F Gd(2\zeta)\right]\right\},\nonumber\\
B(z,t)&=&-A_B\sqrt{\sech{2\zeta}}\,
\exp\left\{i\left[\phi_0+\nu_B Gd(2\zeta)\right]\right\},
\label{FB_exact_soliton}
\end{eqnarray}
where
\begin{eqnarray}
\zeta&=&\frac{t-p(z-z_0)}{\sqrt{1-p^2}},\qquad |p|<1,\nonumber\\
A_F^2&=&\frac{2(1+p)\sqrt{1-p^2}}{3-p^2},\qquad
A_B^2=\frac{2(1-p)\sqrt{1-p^2}}{3-p^2},\nonumber\\
\nu_F&=&\frac{3+4p-p^2}{2(3-p^2)},\qquad
\nu_B=\frac{-3+4p+p^2}{2(3-p^2)}.
\label{FB_exact_parameters}
\end{eqnarray}
Here $z_0$ and $\phi_0$ set the soliton position and common phase. Its centre follows $z=z_0+t/p$, so that $0<|p|<1$ gives the characteristic superluminal peak velocity $|dz/dt|=1/|p|>1$, while the limit $p\rightarrow0$ recovers the spatially uniform temporal homoclinic orbit. In this statement the unit velocity is the characteristic wave speed $v=c/n_r$ used in the normalization of Eq.~(\ref{FB_scaling}); ``superluminal'' therefore means that the trajectory of the intensity maximum, or equivalently the envelope centre, moves faster than this characteristic speed. It does not imply superluminal transport of information or violation of causality. For a finite, truncated excitation the causal forerunner propagates at the front velocity and remains ahead of the reshaped and amplified soliton peak, which cannot overtake it; the apparent faster than light motion is consequently a peak propagation effect produced by temporal amplification and pulse reshaping \cite{segevprl}. For $p\neq0$, the solution has constant integral quantities
\eq{FB_exact_integrals}{\mathcal E_{\rm sol}=\frac{2\pi(1-p^2)}{|p|(3-p^2)},\qquad
\mathcal P_{\rm sol}=p\,\mathcal E_{\rm sol}.}

In our numerical simulations, we initially launch a weak forward-propagating Gaussian pulse,
\eq{small_F_input}{F(z,0)=A_0\exp\left(-\frac{z^2}{2w_0^2}\right),\qquad B(z,0)=0,}
with $A_0=0.1$ and $w_0=5$, centred in the middle of the $k$-gap. The backward component is therefore dynamically generated entirely by the temporal Bragg coupling during the propagation. Since the Gaussian momentum width is smaller than the normalized gap width, most of the input spectrum initially lies inside the fundamental $k$-gap, which is defined by $|\Delta k|<1$. Its different momentum components then experience different rates of parametric amplification, nonlinear phase rotation, and conversion between the forward and backward branches. The unstable part of the pulse initially grows close to the saddle point, but the Kerr terms subsequently saturate the growth and organize the field locally around the homoclinic structure. A sufficiently extended pulse can therefore break into, or nucleate, several well-separated spatiotemporal gap solitons, each with different forward/backward balance in their $F$ and $B$ components, and therefore with different values and signs for their individual parameter $p$.

\begin{figure*}[t]
\centering
\includegraphics[width=0.7\linewidth]{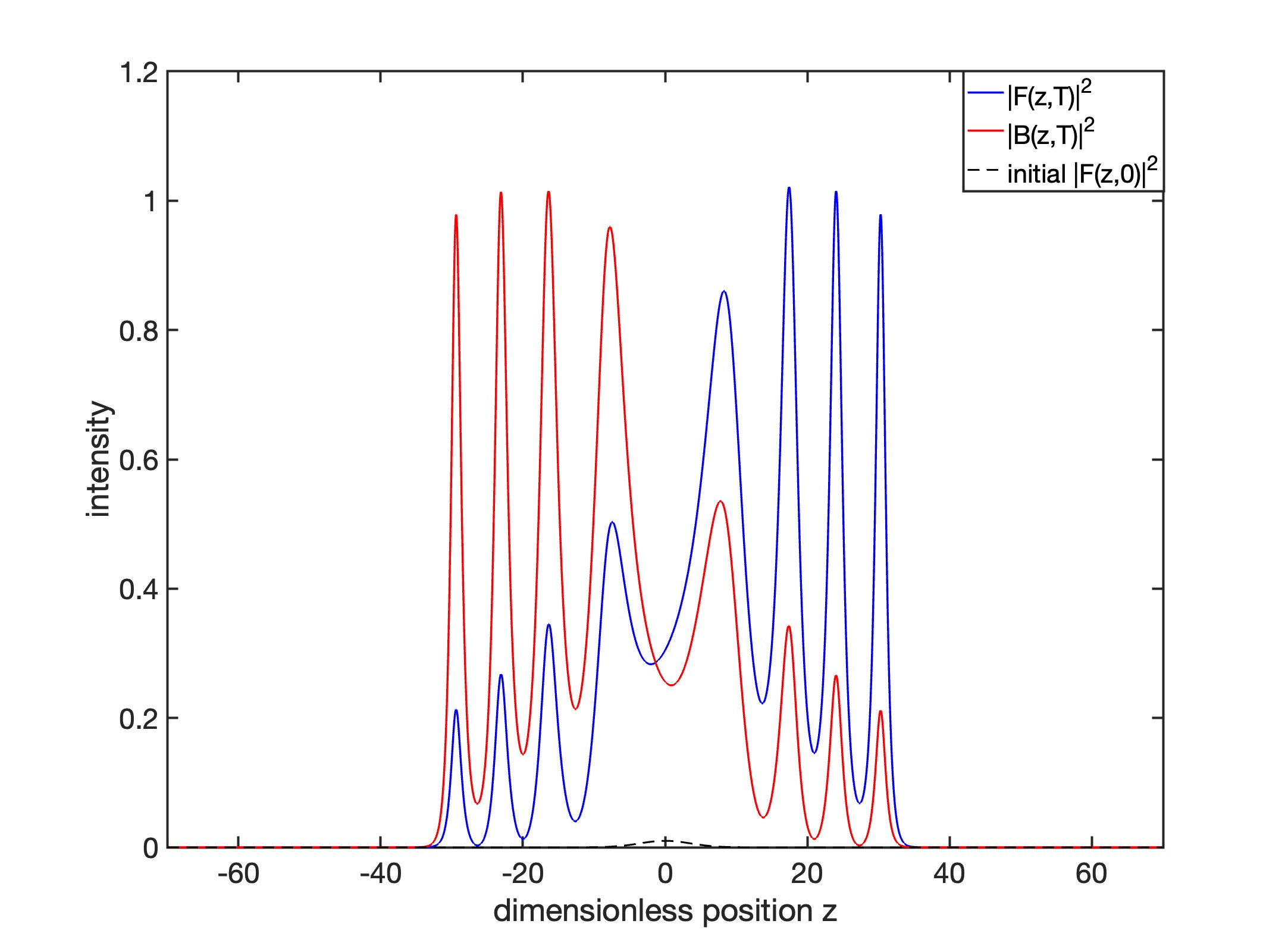}
\caption{Nonlinear fragmentation of the weak forward-only Gaussian input in Eq.~(\ref{small_F_input}), with $A_{0}=0.1$ and $w_{0}=5$, at the final observation time $t=T=15$. The blue and red solid curves show the forward and backward envelope intensities $|F(z,T)|^2$ and $|B(z,T)|^2$, respectively, while the small black dashed curve at the bottom shows the initial forward intensity $|F(z,0)|^2$; we have set a vanishing backward input wave $B(z,0)=0$. The backward field is generated by temporal Bragg coupling, and the Kerr nonlinearity subsequently arrests the local $k$-gap growth, producing a train of sharply localised spatiotemporal gap solitons, described by Eqs. (\ref{FB_exact_soliton}).}
\label{fig:soliton_nucleation}
\end{figure*}

\begin{figure*}[t]
\centering
\includegraphics[width=0.7\linewidth]{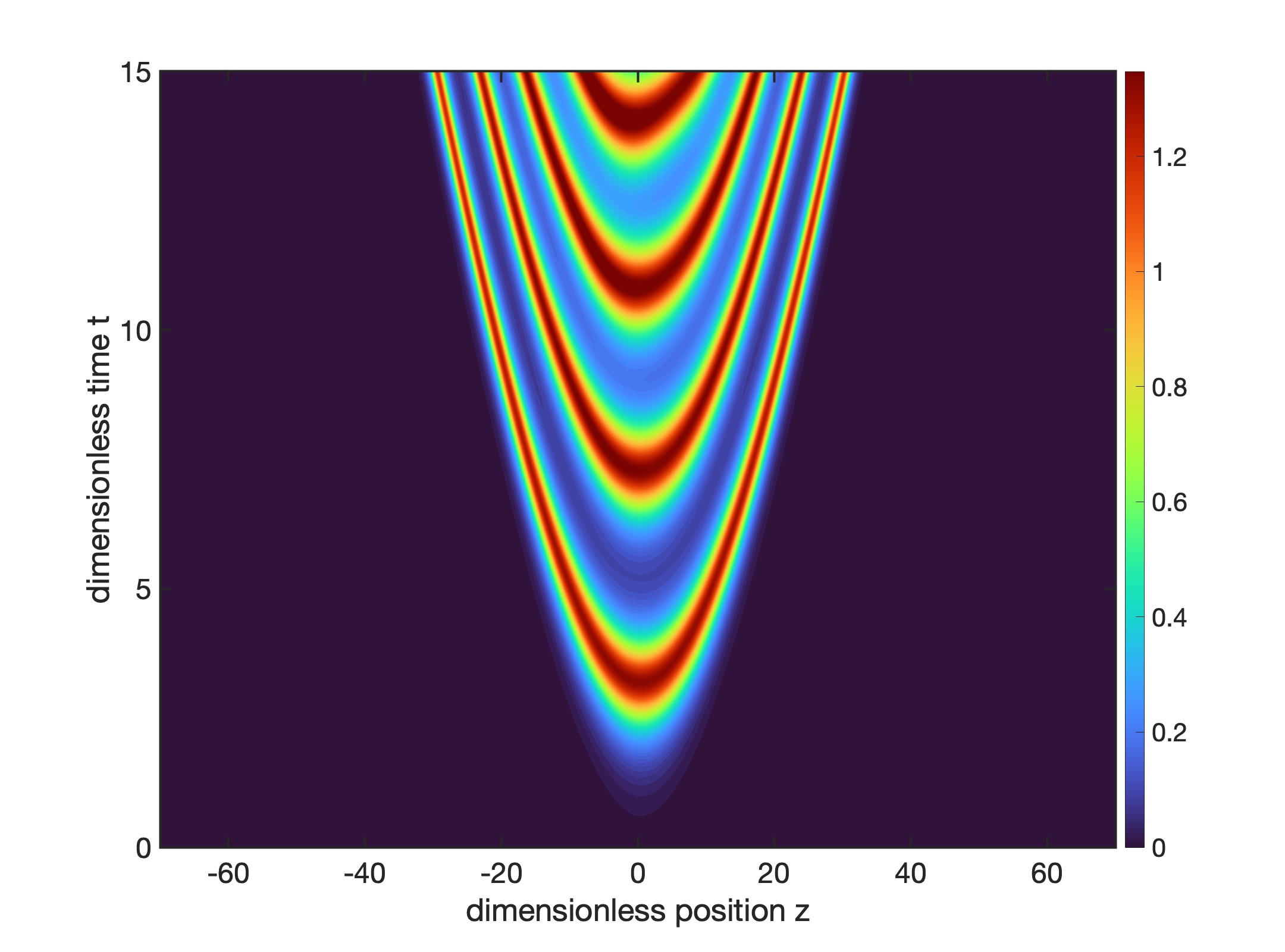}
\caption{Spatiotemporal evolution of the cycle-averaged field intensity $I_{\rm av}(z,t)=|F(z,t)|^2+|B(z,t)|^2$ for the weak forward-only Gaussian input of Eq.~(\ref{small_F_input}). The initially smooth pulse is first amplified by the unstable modes of the fundamental momentum gap while the backward component is generated from zero. Nonlinear phase accumulation subsequently makes the evolution spatially nonuniform and nucleates multiple spatiotemporal gap solitons, with prominent structures propagating in both directions consistently with the conserved momentum $\mathcal P$. Input parameters are identical to those of Fig. \ref{fig:soliton_nucleation}.}
\label{fig:nucleation_map}
\end{figure*}

\begin{figure*}[t]
\centering
\includegraphics[width=0.7\linewidth]{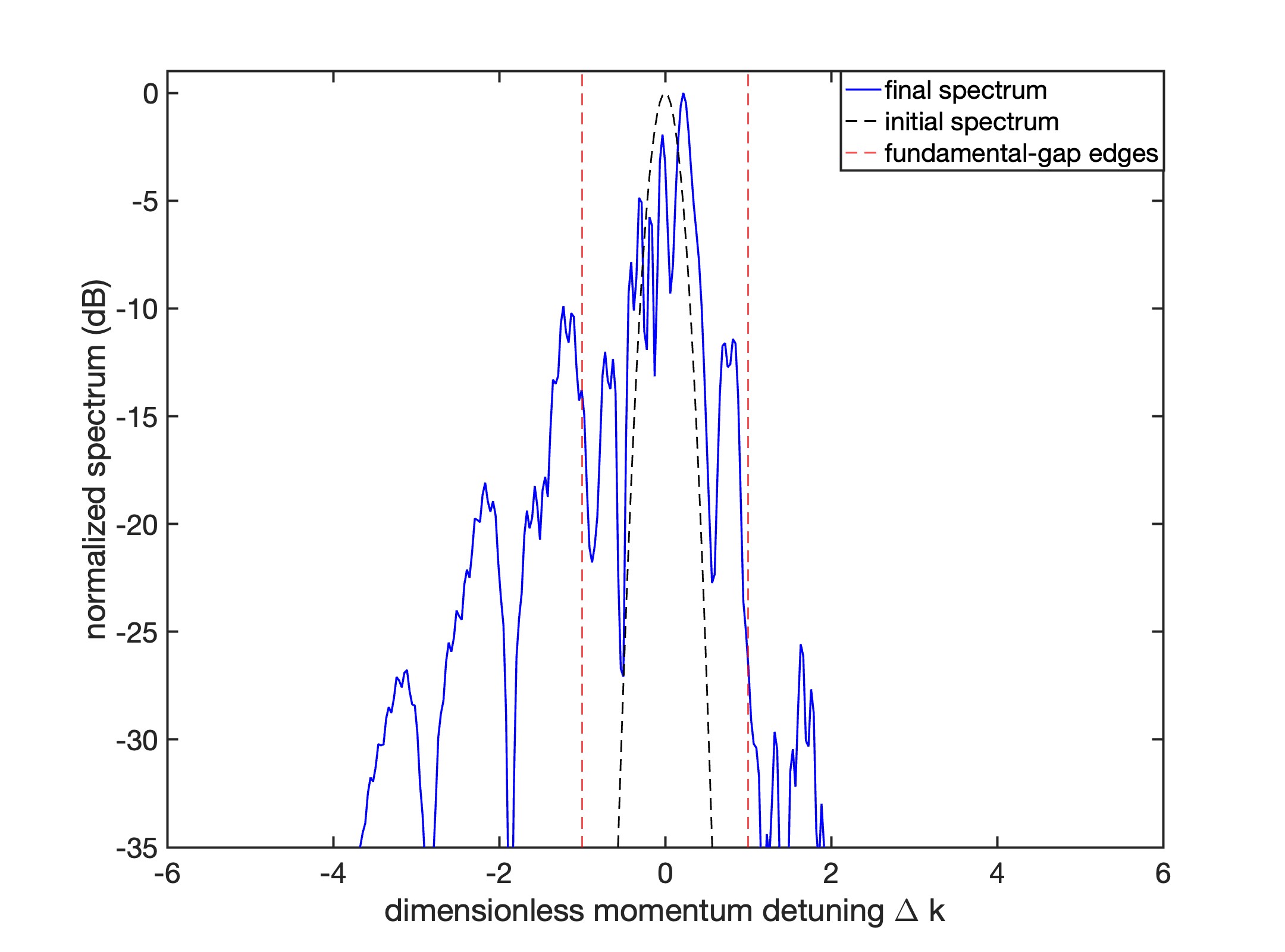}
\caption{Momentum supercontinuum generation from the weak forward-only Gaussian input of Eq.~(\ref{small_F_input}). The plotted quantity is the envelope momentum spectrum defined in Eq.~(\ref{envelope_spectrum}). The black dashed and blue solid curves are the initial and final spectra, respectively. The two red dashed vertical lines mark the edges $\Delta k=\pm1$ of the fundamental momentum gap. The narrow input spectrum lies predominantly inside the gap, whereas the final envelope spectrum goes well beyond both edges as a result of nonlinear localisation and spatiotemporal gap soliton formation. Input parameters are identical to those of Fig. \ref{fig:soliton_nucleation}.}
\label{fig:final_spectrum}
\end{figure*}

\begin{figure*}[t]
\centering
\includegraphics[width=0.7\linewidth]{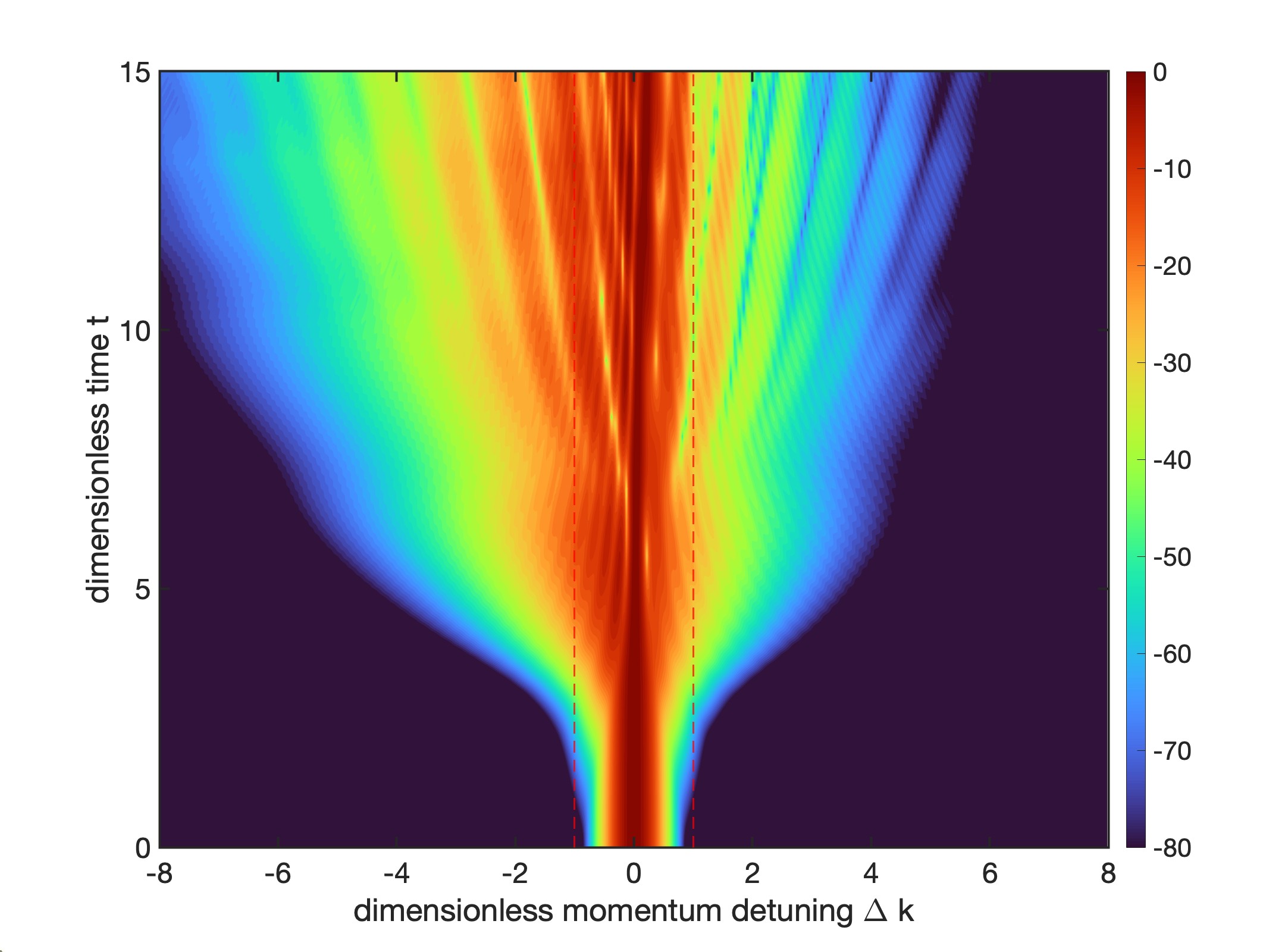}
\caption{Evolution of the momentum spectrum Eq.~(\ref{envelope_spectrum}) for the same weak forward-only Gaussian input as in Fig.~\ref{fig:final_spectrum}. The dashed vertical lines mark the fundamental gap edges $\Delta k=\pm1$. The initially narrow distribution broadens progressively beyond the gap and develops distinct spectral `fingers' associated with the nucleated spatiotemporal gap solitons. Input parameters are identical to those of Fig. \ref{fig:soliton_nucleation}.}
\label{fig:nspectrum_evolution}
\end{figure*}

\begin{figure*}[t]
\centering
\includegraphics[width=0.7\linewidth]{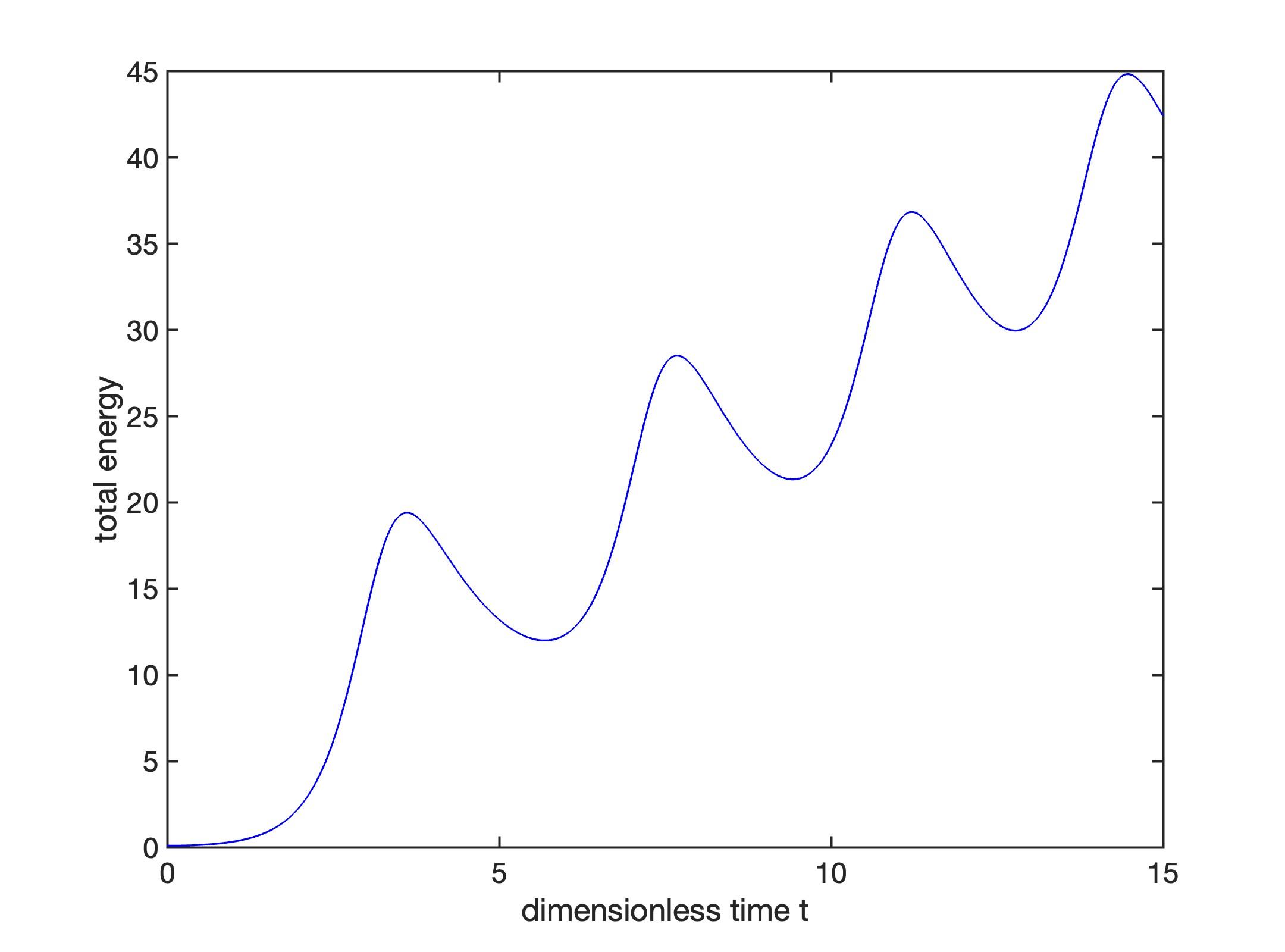}
\caption{Evolution of the total energy integral $\mathcal E(t)=\int (|F|^2+|B|^2)\,dz$ for the weak forward-only Gaussian input of Eq.~(\ref{small_F_input}). Its increase reflects the work performed on the electromagnetic field by the externally driven temporal modulation, consistently with the balance law in Eq.~(\ref{em_balance}); $\mathcal E$ is therefore not a conserved quantity. By contrast, the global momentum $\mathcal P=\int(|F|^2-|B|^2)\,dz$ is conserved by Eq.~(\ref{coupled1}) and is used as a numerical convergence diagnostic.}
\label{fig:total_energy}
\end{figure*}

We integrate Eq.~(\ref{coupled1}) with a symmetric split-step Fourier method, using the exact Fourier space propagator for the linear parts and a fourth-order Runge--Kutta step for the nonlinear part. The numerical results are summarised in Figs.~\ref{fig:soliton_nucleation}--\ref{fig:total_energy}. Figure~\ref{fig:soliton_nucleation} compares the initial small Gaussian seed profile at $t=0$ with the two envelope components at the final evolution time $t=T=15$. The initially smooth, low-amplitude $F$ field first undergoes temporal-Bragg amplification and generates a backward component $B$. As the field grows, Kerr phase accumulation becomes dynamically significant, arrests the local exponential growth, and the pulse begins to nucleate spatiotemporal gap solitons. These structures move towards positive $z$ when their forward component dominates and towards negative $z$ when the backward component dominates. Nucleation continues throughout the simulated time window; in a realistic implementation, losses, finite modulation duration, pump depletion, and other saturation mechanisms will eventually limit this process.

The corresponding spacetime map in Fig.~\ref{fig:nucleation_map} shows the dynamical evolution of this nucleation process: the weak pulse is first amplified inside the fundamental gap, after which nonlinear phase shifts and forward--backward conversion break the extended excitation into distinct spatiotemporal gap solitons. Prominent structures propagate in both directions, consistently with conservation of the global momentum $\mathcal{P}$.

To quantify the momentum redistribution due to nucleation, in Figs.~\ref{fig:final_spectrum} and \ref{fig:nspectrum_evolution} we consider the momentum spectrum of the slowly varying envelopes rather than the carrier-resolved Fourier spectrum of the real displacement field. We define
\eq{envelope_spectrum}{S_{\rm env}(\Delta k,t)\equiv |\widetilde F(\Delta k,t)|^2+|\widetilde B(\Delta k,t)|^2,}
where $\widetilde F(\Delta k,t)=\int F(z,t)e^{-i\Delta k z}dz$ and analogously for $\widetilde B$. This quantity directly measures the distribution of the coupled forward and backward envelopes in momentum detuning and removes the rapidly oscillating carrier phase interference that would be present in an instantaneous reconstruction of $D$. As shown in Fig.~\ref{fig:final_spectrum}, the initial spectrum is concentrated mainly between the fundamental gap edges $\Delta k=\pm1$, whereas the final spectrum extends far outside this interval. This broadening is generated by the progressively finer spatial structure of the nucleated peaks: localisation requires a wide range of Fourier components, while the unequal positions and phases of the spatiotemporal gap solitons superimpose an interference modulation on the broad spectral envelope. We refer to this phenomenon as {\em momentum supercontinuum generation}, and it is one of the main results of this paper. In Fig.~\ref{fig:nspectrum_evolution} we show the evolution of $S_{\rm env}(\Delta k,t)$ during the propagation. The spectrum undergoes a significant broadening that extends outside the $k$-gap, while the distinct solitonic `fingers' on both sides of the spectrum are clearly visible.

We show the energy evolution of the system in Fig.~\ref{fig:total_energy}. According to Eq.~(\ref{em_balance}), the temporal modulation can either supply energy to the field or extract energy from it, depending on the relative phase of the forward and backward components. For the forward-only input pulses considered here, the net exchange is positive over the simulated interval: $\mathcal{E}(t)$ exhibits an overall increase, despite temporary depletion oscillations. At the same time, $\mathcal{P}$ is perfectly conserved and provides a stringent numerical check of the propagation.

\begin{figure*}[t]
\centering
\includegraphics[width=0.7\linewidth]{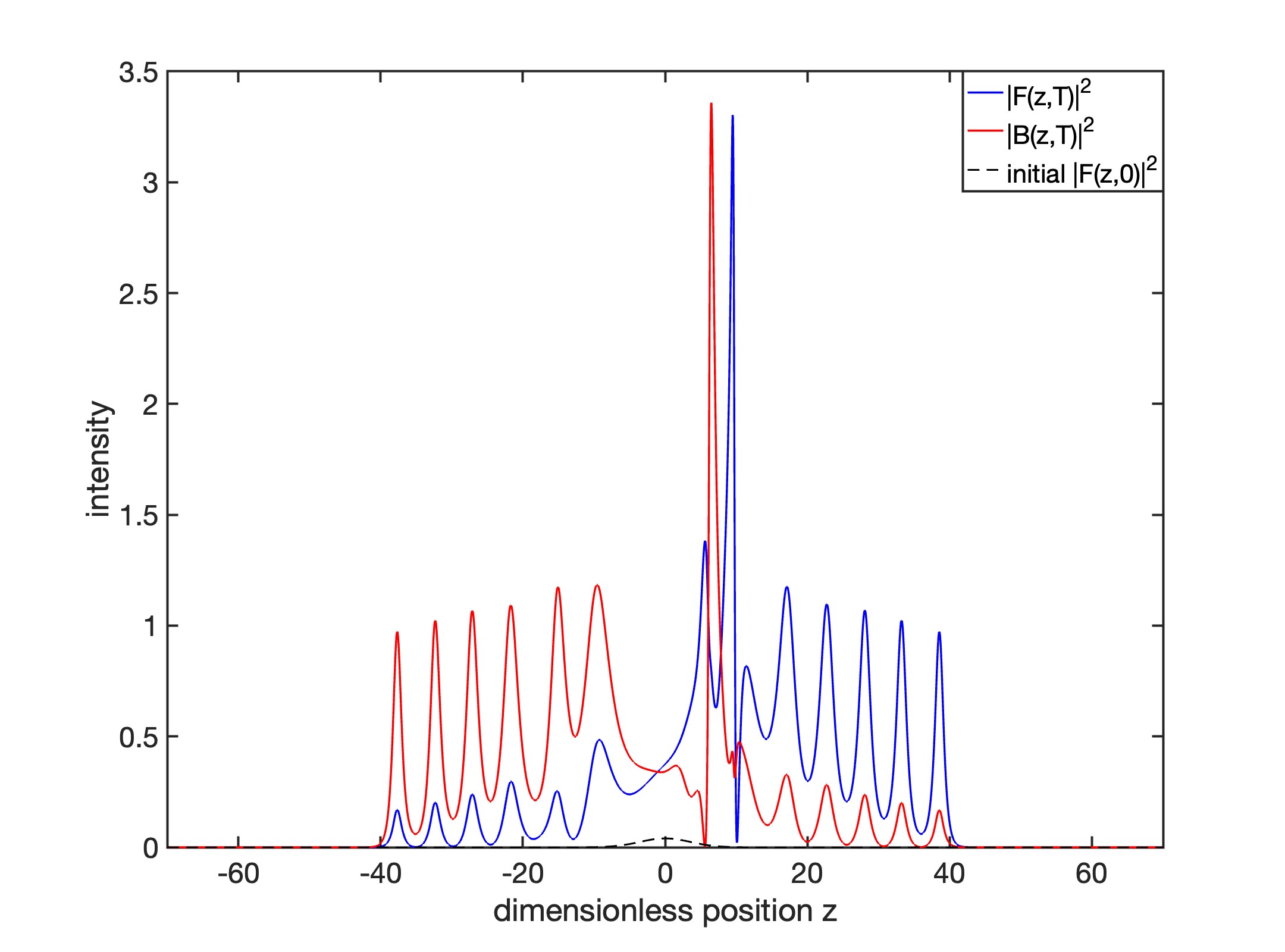}
\caption{Final forward and backward envelope intensities for the stronger forward-only Gaussian input of Eq.~(\ref{small_F_input}), with $A_0=0.2$, $w_0=5$, and $B(z,0)=0$, at $t=T=20$. The blue and red solid curves show $|F(z,T)|^2$ and $|B(z,T)|^2$, respectively, while the black dashed curve shows the initial forward intensity $|F(z,0)|^2$. Compared with the weaker input case in Fig.~\ref{fig:soliton_nucleation}, the soliton train develops a small number of exceptionally intense peaks, signalling the extreme nonlinear localisation event discussed in the text.}
\label{fig:soliton_nucleation2}
\end{figure*}

\begin{figure*}[t]
\centering
\includegraphics[width=0.7\linewidth]{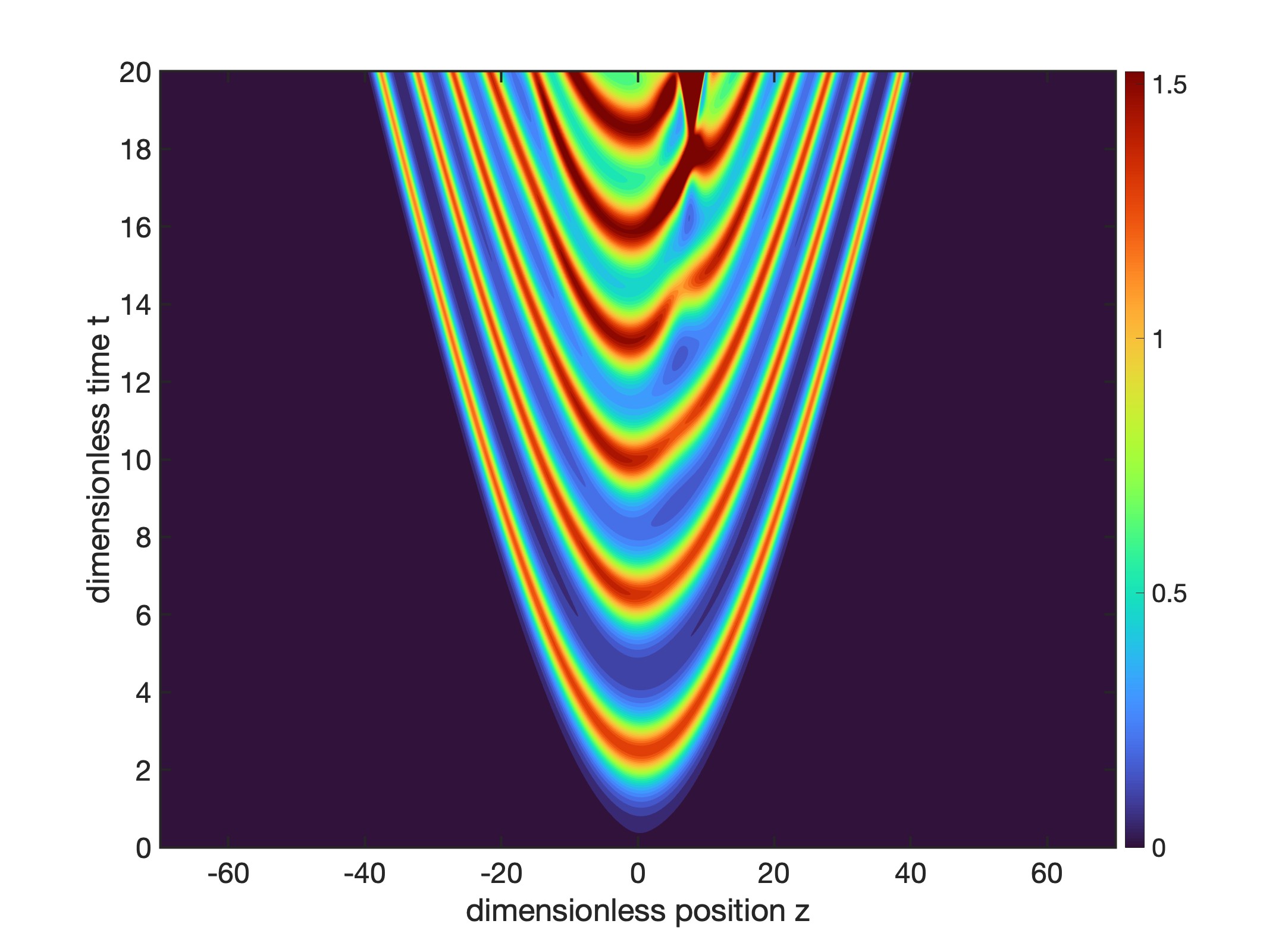}
\caption{Spatiotemporal evolution of the cycle-averaged envelope intensity $I_{\rm av}(z,t)=|F(z,t)|^2+|B(z,t)|^2$ for the stronger input $A_0=0.2$ and $w_0=5$. As in Fig.~\ref{fig:nucleation_map}, temporal Bragg amplification and Kerr saturation generate multiple spatiotemporal gap solitons. At late times, around $t\simeq18$, a strongly localised, high-intensity event emerges from the nonlinear soliton dynamics; its final spatial signature is visible in Fig.~\ref{fig:soliton_nucleation2}.}
\label{fig:nucleation_map2}
\end{figure*}

Doubling the input field amplitude from $A_{0}=0.1$ to $A_{0}=0.2$ (and hence increasing the input peak intensity by a factor of four), while maintaining the same width $w_{0}=5$, leads to a qualitatively stronger localisation regime. In Fig. \ref{fig:soliton_nucleation2} we show the spatial structure of the field at a propagation time $T=20$. It is very similar to Fig. \ref{fig:soliton_nucleation}, but shows the presence of a couple of peaks with amplitude much larger than the others. The moment when this `extreme event' is produced is clearly visible in the contour map of the propagation, Fig. \ref{fig:nucleation_map2}, around $t\simeq 18$.
Related `breathing' $k$-gap events have recently been reported theoretically by Zhang \etal, where a spatially localised perturbation of an unstable nonlinear $k$-gap soliton background nucleates a transient, strongly localised spatiotemporal event \cite{zhang_breathing}. The present dynamics is generated instead from a single finite-width forward seed with $B(z,0)=0$, and here we emphasize in particular the abrupt $k$-spectrum broadening associated with the extreme event.

\begin{figure*}[t]
\centering
\includegraphics[width=0.7\linewidth]{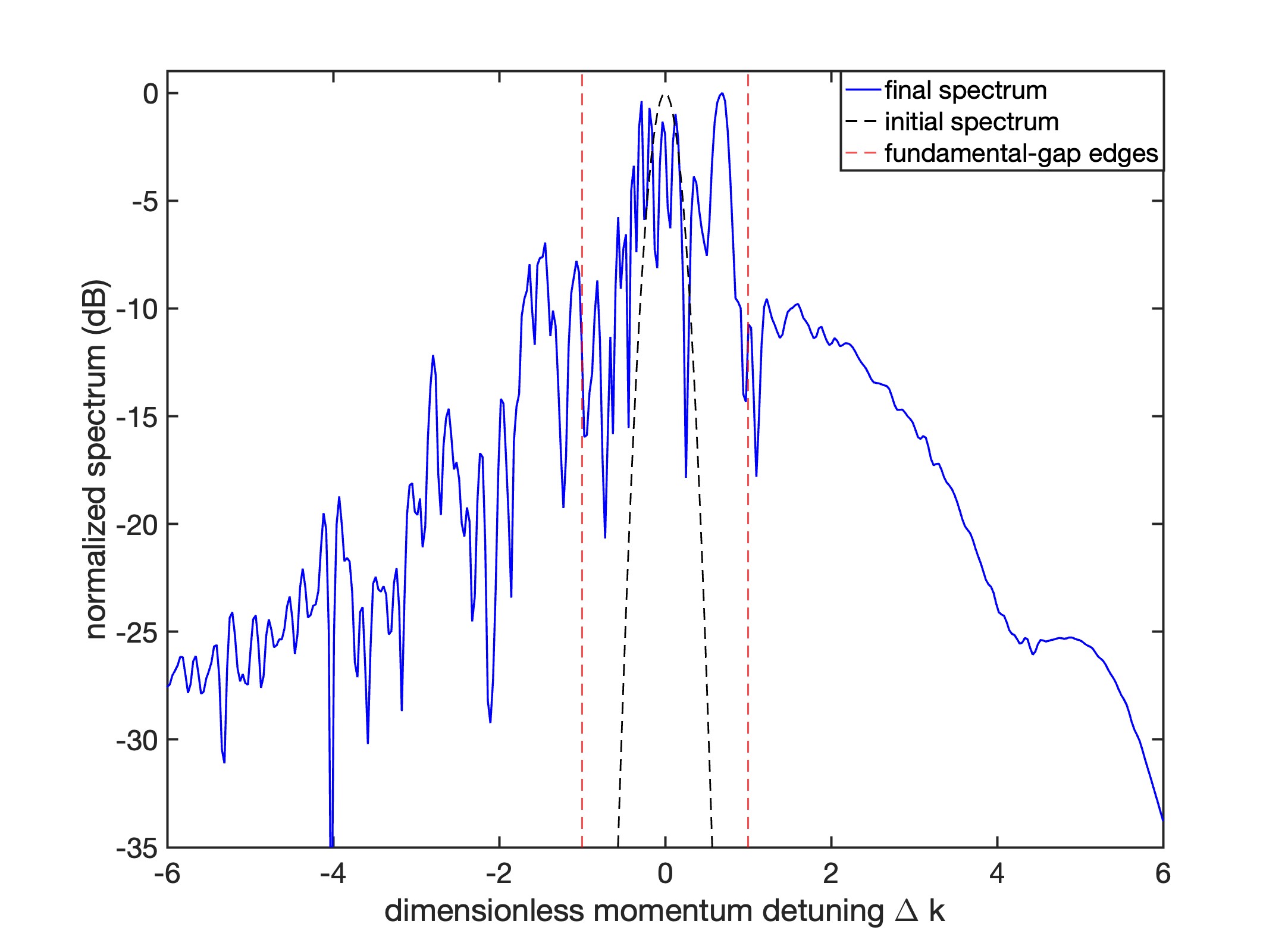}
\caption{Momentum supercontinuum spectrum for the stronger forward-only Gaussian input with $A_0=0.2$ and $w_0=5$ at $t=T=20$. As in Fig.~\ref{fig:final_spectrum}, the plotted quantity is the envelope spectrum defined in Eq.~(\ref{envelope_spectrum}), with the initial and final spectra shown by the black dashed and blue solid curves, respectively. The red dashed vertical lines indicate the fundamental gap edges $\Delta k=\pm1$. Relative to the weak input case, the final spectrum develops a pronounced broad tail, particularly for positive $\Delta k$, associated with the extreme localisation event shown in Figs.~\ref{fig:soliton_nucleation2} and \ref{fig:nucleation_map2}.}
\label{fig:final_spectrum2}
\end{figure*}

\begin{figure*}[t]
\centering
\includegraphics[width=0.7\linewidth]{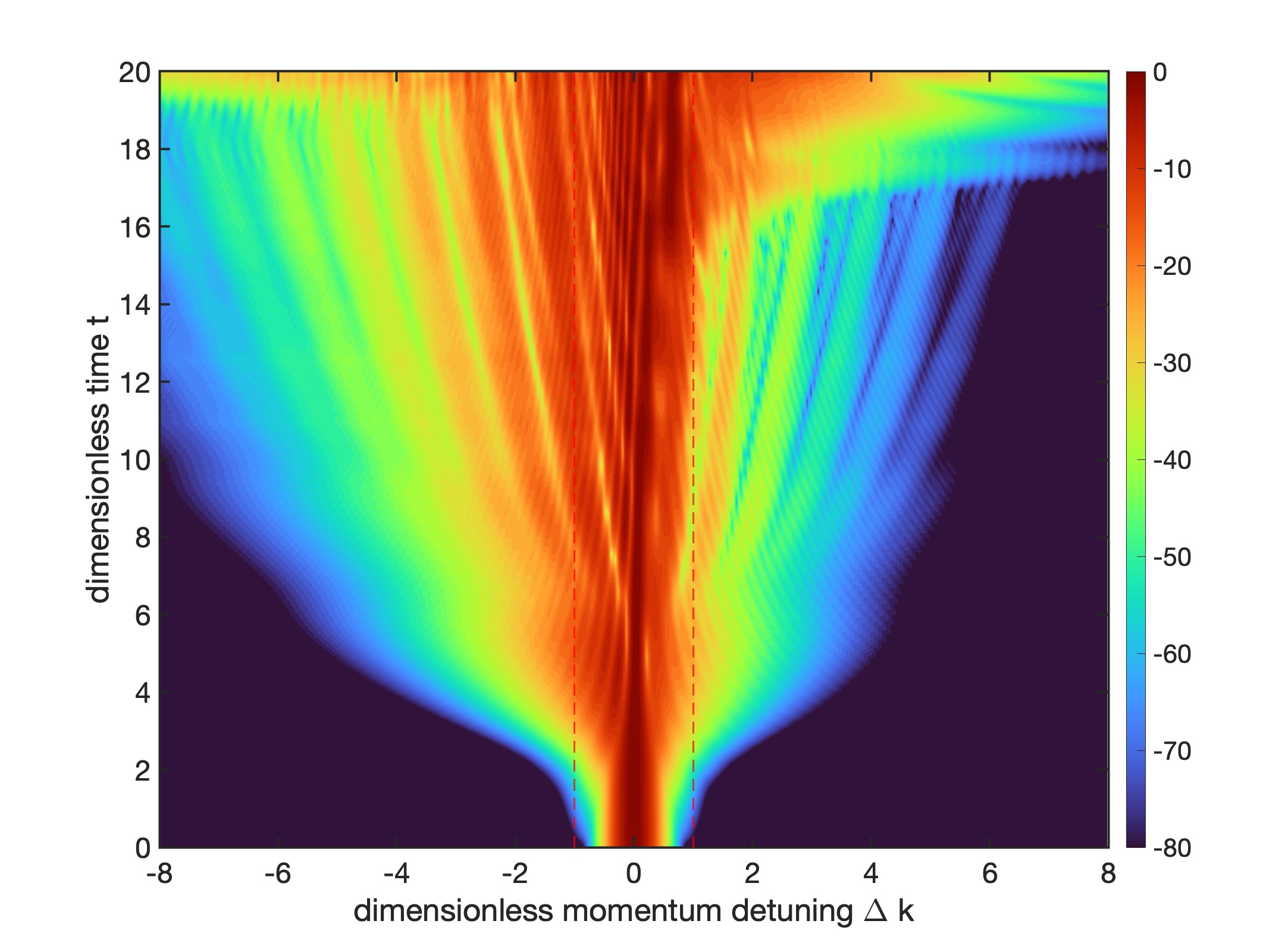}
\caption{Evolution of the envelope momentum spectrum of Eq.~(\ref{envelope_spectrum}) for the stronger input $A_0=0.2$ and $w_0=5$. The dashed vertical lines mark the fundamental gap edges $\Delta k=\pm1$. The initially narrow distribution broadens progressively as spatiotemporal gap solitons are nucleated. Near $t\simeq18$, coincident with the extreme localisation event in Fig.~\ref{fig:nucleation_map2}, the spectrum undergoes an abrupt additional expansion well beyond the $k$-gap, producing the broad continuum observed at the final time in Fig.~\ref{fig:final_spectrum2}.}
\label{fig:spectrum_evolution2}
\end{figure*}

The final spectrum is shown in Fig.~\ref{fig:final_spectrum2}. Relative to Fig.~\ref{fig:final_spectrum}, a pronounced broad tail develops, particularly at positive $\Delta k$. Its onset is temporally associated with the extreme localisation event visible in Fig.~\ref{fig:nucleation_map2}, and the time-resolved spectrum in Fig.~\ref{fig:spectrum_evolution2} shows the corresponding abrupt additional broadening. This correlation indicates that strong transient localisation can substantially extend the momentum continuum beyond the width produced by the weaker soliton-nucleation dynamics.

The momentum supercontinuum considered here is distinct from the conventional frequency supercontinuum generated in optical fibres. In a stationary fibre, dispersion and Kerr nonlinearity redistribute optical frequency during spatial propagation, whereas in a PTC the externally driven temporal modulation can exchange energy and frequency with the field while spatial homogeneity preserves the global momentum $\mathcal{P}$. In a realistic implementation, loss, finite modulation duration, pump depletion, material dispersion, and saturation of the nonlinear response will eventually limit the dynamics. These effects will be essential for quantitative experimental predictions, but the mechanism identified here remains the same: $k$-gap amplification of a finite-width excitation is converted by the Kerr response into localised nonlinear structures whose progressively finer spatial features generate a broad momentum continuum. This process is still largely unexplored, and we have no doubt that it will generate new exciting physics in the near future.

\clearpage
\section{Conclusions and future perspectives}

In conclusion, we have developed a unified reduced description of nonlinear wave dynamics near the first momentum gap of a shallow sinusoidally modulated PTC, starting from Maxwell's equations for the displacement field. For spatially monochromatic excitation, the problem reduces to a nonlinear Mathieu equation and, near the first gap, to an autonomous Hamiltonian system. This formulation yields the temporal homoclinic gap solitons throughout the first instability tongue, their continuous phase branch, the complete phase-space topology of the reduced system, and the point of maximum linear Floquet gain.

Exactly at the first temporal Bragg resonance, retaining the spatial dependence leads to coupled forward and backward envelope equations. The temporal modulation can exchange energy with the optical field, while the global momentum $\mathcal P$ remains conserved. A finite forward-only input is amplified and reshaped by the Kerr nonlinearity into multiple spatiotemporal gap solitons, whose localisation produces a broad momentum distribution extending well outside the original $k$-gap. This provides the nonlinear PTC mechanism for momentum-supercontinuum generation identified here; for stronger excitation, transient extreme localisation events are correlated with a pronounced, abrupt additional spectral broadening.

Important next steps would be to include material dispersion, loss, finite switching and modulation duration, pump depletion, and delayed or saturating nonlinear response, and to derive the coupled-mode coefficients for specific experimental platforms. Quantifying the stability and interaction laws of the nucleated solitons will also be essential for assessing the momentum supercontinuum generation in current and future PTC implementations.

\section*{Acknowledgements}
The author would like to thank Prof M. Ferrera, Dr S. Stengel and Dr W. Jaffray (all from Heriot-Watt University, Edinburgh, UK) for past useful discussions.

\end{document}